\documentclass[12pt,a4paper,final]{iopart}
\usepackage{iopams,graphicx}
\usepackage[breaklinks=true,colorlinks=true,linkcolor=blue,urlcolor=blue,citecolor=blue]{hyperref}
\usepackage{color}

\def\bra#1{\mathinner{\langle{#1}|}}
\def\ket#1{\mathinner{|{#1}\rangle}}
\def\braket#1{\mathinner{\langle{#1}\rangle}}

{\catcode`\|=\active 
  \gdef\Braket#1{\begingroup
\mathcode`\|32768\let|\BraVert\left<{#1}\right>\endgroup}}
\def\BraVert{\egroup\,\mid\,\bgroup}

\newcommand{\fisher}{\mathcal{F}}

\definecolor{Blue}{rgb}{0,0,1}
\definecolor{Red}{rgb}{1,0,0}

\begin{document}

\title{Coherent measurements in quantum metrology}

\author{K. Micadei$^1$, D. A. Rowlands$^2$, F. A. Pollock$^{3,5}$, L. C. C\'{e}leri$^4$, R. M. Serra$^1$, and K. Modi$^5$}
\address{$^1$Centro de Ci\^{e}ncias Naturais e Humanas, Universidade Federal do ABC, Rua Santa Ad\'{e}lia 166, 09210-170, Santo Andr\'{e}, S\~{a}o Paulo, Brazil}

\address{$^2$Cavendish Laboratory, University of Cambridge, JJ Thomson Avenue, Cambridge, CB3 0HE, United Kingdom}

\address{$^3$Department of Physics, University of Oxford, Clarendon Laboratory, Oxford, OX1 3PU, United Kingdom}

\address{$^4$Instituto de F\'{i}sica, Universidade Federal de Goi\'{a}s, 74.001-970, Goi\^{a}nia, Goi\'{a}s, Brazil}

\address{$^5$School of Physics, Monash University, Victoria 3800, Australia}
\ead{kavan.modi@monash.edu}

\begin{abstract}
It is well known that a quantum correlated probe can yield better precision in estimating an unknown parameter than classically possible. However, how such a quantum probe should be measured remains somewhat elusive. We examine the role of measurements in quantum metrology by considering two types of readout strategies: \emph{coherent}, where all probes are measured simultaneously in an entangled basis; and \emph{adaptive}, where probes are measured sequentially, with each measurement one way conditioned on the prior outcomes. Here we firstly show that for classically correlated probes the two readout strategies yield the same precision. Secondly, we construct an example of a noisy multipartite quantum system where coherent readout yields considerably better precision than adaptive readout. This highlights a fundamental difference between classical and quantum parameter estimation. From the practical point of view, our findings are relevant for the optimal design of precision-measurement quantum devices.
\end{abstract}

\vspace{2pc}
\submitto{\NJP}
%\maketitle

%%%%%%%%%%%%%%%%%%%%%%%%%%%%%%%%%%%%%%%%%%%%%%%%%%%%%%%%%%%%
%%%%%%%%%%%%%%%%%%%%%%%%%%%%%%%%%%%%%%%%%%%%%%%%%%%%%%%%%%%%
\section{Introduction}
%%%%%%%%%%%%%%%%%%%%%%%%%%%%%%%%%%%%%%%%%%%%%%%%%%%%%%%%%%%%
%%%%%%%%%%%%%%%%%%%%%%%%%%%%%%%%%%%%%%%%%%%%%%%%%%%%%%%%%%%%
Quantum mechanical systems can be used to outperform classical ones in information processing~\cite{Nielsen}. Quantum correlations can be employed to beat the shot-noise (standard) limit in metrology protocols. Such parameter estimation methods are crucial for both theoretical advances and the development of technologies. However, almost all quantum technologies operate with some level of noise and how quantum enhancement fares in the presence of noise is still unclear. Another intrinsic quantum feature is the measurement process, which, in general, disturbs the system being measured and the outcome depends on a basis choice. Here, we explore this distinction to uncover a difference between coherent and adaptive readout strategies in a quantum metrology protocol in the presence of noise. 

A general framework to estimate a parameter involves a suitable probe and an interaction that physically manifests the parameter. The probe, initially in state $\varrho$, acquires some information about the parameter, $\phi$, yielding the encoded state $\varrho_{\phi}$, which is then read out by some convenient strategy. The estimation process depends on how much information about the parameter is encoded in the probe. The precision of the estimation protocol, which can be saturated in a large number of trials, is limited by the Cram\'{e}r-Rao relation~\cite{Cramer, Rao, Kullback}, in which the root mean square error, $\Delta \phi$, associated with an unbiased estimator, is bounded by the Fisher information~\cite{Fisher}, $\fisher$, as $\Delta \phi \geq 1/\sqrt{\fisher}$. Fisher information is a key concept in metrology and gives us knowledge about the effectiveness of a parameter estimation protocol. Measuring the encoded probe state affords the probability distribution $p_\phi(x)$, from which $\fisher$ can be computed directly from its definition, $\fisher = \sum_x p_{\phi}(x) [\partial_\phi \ln p_{\phi}(x) ]^2$ (for discrete outcomes $x$). The quantum Fisher information (QFI) is defined as the maximum of $\fisher$ over all possible measurements, and to attain it we need to employ a readout procedure that yields an appropriate distribution $p_\phi(x)$.

A valuable ingredient in parameter estimation is the use of correlated probes, which can be employed to improve the estimation. For instance, in the quantum case nonclassical correlations offer considerable advantages in quantum metrology~\cite{Maccone, Maccone1, Maccone2} even in noisy scenarios~\cite{Modi, Escher1, Escher2}. Such studies are tractable because QFI can be computed using only the initial probe state and the generator of the encoding~\cite{Caves, Caves2, Luo0, Paris, Hayashi}. Regarding the readout strategy in quantum metrology, in Ref.~\cite{Ballester1} it was found that entangled measurements were important for the optimal estimation of an unknown unitary. On the other hand in~\cite{Ballester2}, it is argued that, for a restricted class of states, entangled measurements did not provide an advantage for phase estimation in the studied cases. Very recently, an experiment employed a joint measurement of phase and phase diffusion as an optimum measurement scheme~\cite{Vidrighin}. In the present article we address the need for correlations in the readout procedure for estimating a single real parameter. That is, the need for entangling measurements to produce a suitable $p_\phi(x)$. Specifically, we show that for some correlated probe states a coherent quantum readout procedure yields a better estimate for the parameter of interest than an adaptive classical readout procedure. We begin by defining coherent and adaptive measurements below. However, for simplicity's sake we only consider a bipartite probe for now and address the general multipartite case at the end of the article.

%%%%%%%%%%%%%%%%%%%%%%%%%%%%%%%%%%%%%%%%%%%%%%%%%%%%%%%%%%%%
%%%%%%%%%%%%%%%%%%%%%%%%%%%%%%%%%%%%%%%%%%%%%%%%%%%%%%%%%%%%
\section{Coherent and adaptive measurements}
%%%%%%%%%%%%%%%%%%%%%%%%%%%%%%%%%%%%%%%%%%%%%%%%%%%%%%%%%%%%
%%%%%%%%%%%%%%%%%%%%%%%%%%%%%%%%%%%%%%%%%%%%%%%%%%%%%%%%%%%%
Let us consider the following game between three characters: Alice, Bob, and Charlie. They receive the same bipartite encoded probe $\varrho_{\phi}$. However, Charlie has access to the whole state, while Alice and Bob have access only to the local partitions $A$ and $B$, respectively. We suppose that Charlie has an apparatus able to perform joint measurements on the whole state yielding a bipartite probability distribution $p_{\phi}(a,b)$. The precision that Charlie can attain about $\phi$ is bounded by Fisher information
\begin{equation}\label{jointFisher}
\fisher(A,B) = \sum_{a,b} p_{\phi}(a,b) \left[{\partial_\phi} \ln p_{\phi}(a,b)\right]^2,
\end{equation}
with $a$ and $b$ being the outcomes associated with partitions $A$ and $B$, respectively. 

Alice and Bob are allowed to communicate and perform any operations on their own partitions. To measure $\phi$ with optimal precision Alice and Bob can employ a one-way adaptive strategy. First, Bob performs a suitable measurement on his partition and observes outcome $b$ with probability $q_\phi(b)$. He then communicates his result to Alice, and, based on that Alice performs a measurement on her partition observing outcome $a$ with probability $q_\phi(a|b)$. Putting their outcomes together, the precision with which Alice and Bob can attain $\phi$ is bounded by the Fisher information of the joint probability distribution $q_\phi(a,b) = q_\phi(b) q_\phi(a|b)$. That is, replace $p_{\phi}(a,b)$ with $q_{\phi}(a,b)$ in Eq.~(\ref{jointFisher}). Alternatively, the precision with which Bob can attain $\phi$ is bounded by Bob's Fisher information 
\begin{equation}
\fisher(B)= \sum_{b} q_{\phi}(b) 
\left[{\partial_\phi} \ln q_{\phi}(b) \right]^2.
\end{equation}
Similarly, Alice's precision is bounded by the conditional Fisher information for the conditional distribution $q_{\phi}(a|b) = q_\phi(a,b) / q_\phi(b)$:
\begin{equation}
\fisher(A|B)\equiv \sum_{b} q_{\phi}(b) \fisher(A|B=b),
\end{equation}
with $\fisher(A|B=b) \equiv \sum_a q_{\phi}(a|b) \left[ \partial_\phi \ln q_{\phi}(a|b)\right]^2$ being the Fisher information for Alice conditioned on Bob's outcome $b$. Together their precision of $\phi$ is bounded by $\fisher(B)+\fisher(A|B)$. In general Alice and Bob may use ancillas to implement a many-round readout. In Eq.~(\ref{manyrounds}) of \ref{mult} we give a generalised result for such a readout strategy.

%%%%%%%%%%%%%%%%%%%%%%%%%%%%%%%%%%%%%%%%%%%%%%%%%%%%%%%%%%%%
\begin{figure}
\begin{center}
\includegraphics[scale=.6]{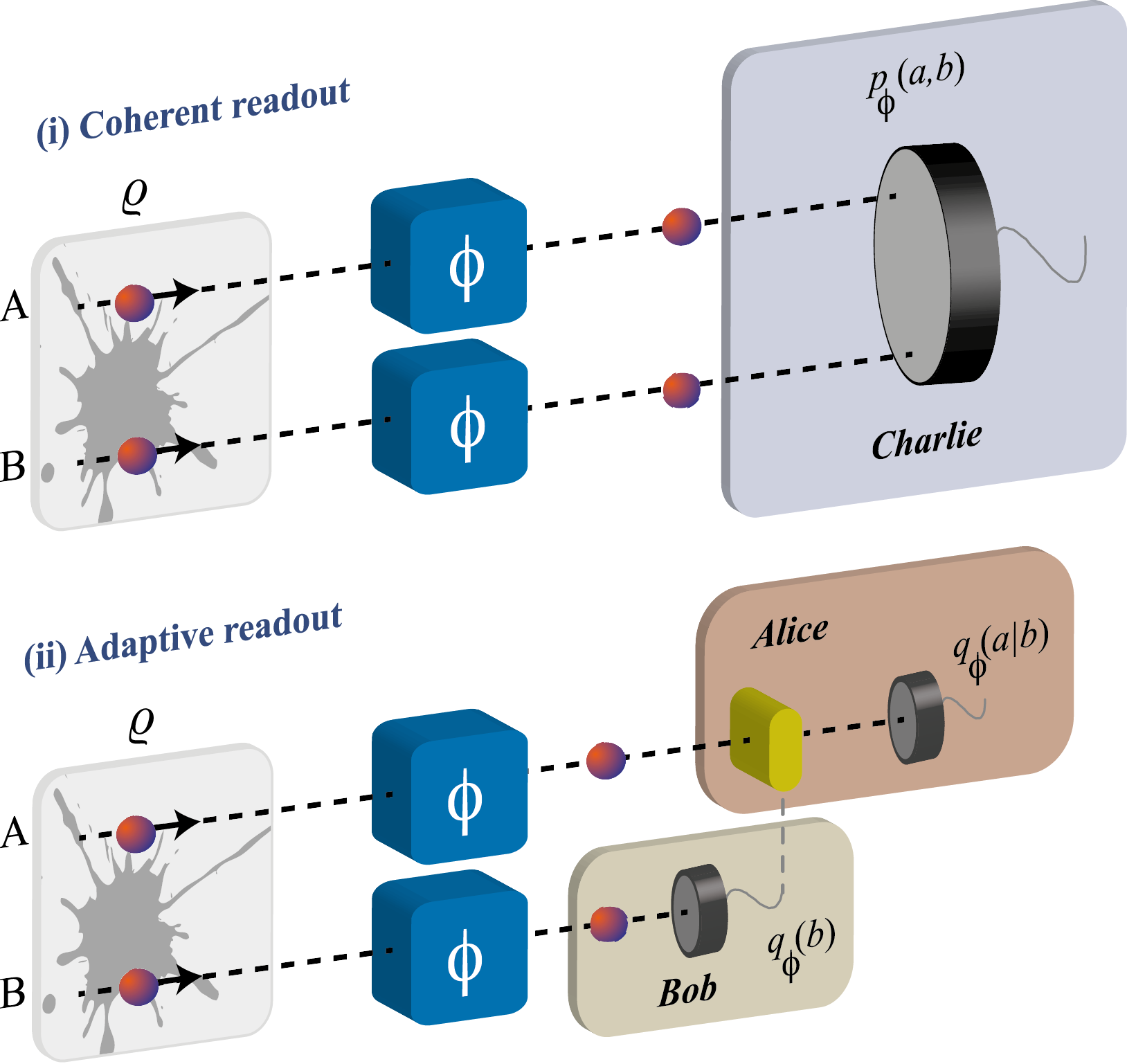}
\end{center}
\caption{\label{StratFig}Two readout strategies for parameter estimation. In (i) the coherent readout process consists of applying an entangling measurement. An entangling measurement can be achieved by performing an entangling unitary followed by measuring both partitions locally. In (ii) Bob measures his partition and observes outcome $b$, which he communicates to Alice. Next, Alice measures her partition, conditioned on Bob's outcome, and observes outcome $a|b$. For classical metrology, (i) yields the left-hand side of Eq.~(\ref{Eq1}) and (ii) yields the right-hand side of Eq.~(\ref{Eq1}). The key ingredient for coherent readout is the ability to perform an entangling measurement.}
\end{figure}
%%%%%%%%%%%%%%%%%%%%%%%%%%%%%%%%%%%%%%%%%%%%%%%%%%%%%%%%%%%%

We are now ready to formally state our problem: we consider a challenge by Charlie to Alice and Bob to attain the same precision for $\phi$ as he does using the coherent strategy. When Alice and Bob are able to meet Charlie's challenge, no quantum resources are necessary for the readout. We will show below that this is indeed the case in classical metrology and even $N00N$ or the equivalent $N-$qubit Greenberger-Horne-Zeilinger (NGHZ) state (both to be defined below) metrology in the absence of noise. We then show that with the introduction of noise there is a gap in the achievable precision between the two readout strategies. We have graphically depicted the two readout strategies in Fig.~\ref{StratFig}.

%%%%%%%%%%%%%%%%%%%%%%%%%%%%%%%%%%%%%%%%%%%%%%%%%%%%%%%%%%%%
\subsection{Classical metrology} 
%%%%%%%%%%%%%%%%%%%%%%%%%%%%%%%%%%%%%%%%%%%%%%%%%%%%%%%%%%%%
In classical metrology, the state of the probe is simply a probability distribution $p_0(a,b)$ and the encoded state is also a probability distribution $p_\phi(a,b)$. Classical probabilities satisfy 
\begin{equation}\label{Bayes}
p_{\phi}(a,b) = p_{\phi}(b) p_{\phi}(a|b).
\end{equation}
The left- and right-hand sides of Eq.~(\ref{Bayes}) correspond to Charlie's coherent readout, and Alice and Bob's adaptive readout, respectively. This means that for classical metrology we have the following well-known additivity relation for Fisher information~\cite{Kullback}
\begin{equation}\label{Eq1}
\fisher(A,B) = \fisher(B) +\fisher(A|B).
\end{equation}
The implication is that Alice and Bob are able to attain the same precision as Charlie for $\phi$ and thus meet Charlie's challenge. We have given a full derivation of Eq.~(\ref{Eq1}), including the multipartite case, in~\ref{App1}.

%%%%%%%%%%%%%%%%%%%%%%%%%%%%%%%%%%%%%%%%%%%%%%%%%%%%%%%%%%%%
\subsection{Quantum metrology}
%%%%%%%%%%%%%%%%%%%%%%%%%%%%%%%%%%%%%%%%%%%%%%%%%%%%%%%%%%%%
In the quantum setting, the  discrete probability distribution $p_{\phi}(x)$ is now obtained from a \emph{positive operator-valued measure} (POVM) $\mathbf{\Pi}$ with elements $\{\Pi_{x}\}$ acting on the encoded quantum state $\varrho_{\phi}$: $p_{\phi}(x)=\tr \left[\varrho_{\phi} \Pi_{x} \right]$ with $\sum_{x} \Pi_{x}=\mathbb{I}$. Remarkably, there is a simple formula that yields the optimal Fisher information over all POVMs~\cite{Caves}:
\begin{equation}\label{qFisher}
\max_{\mathbf{\Pi}} \fisher(\varrho_\phi)
= 2 \sum_{ij} \frac{(\lambda_i - \lambda_j)^2}{\lambda_i + \lambda_j} \left\vert \braket{\psi_i \vert H \vert \psi_j} \right\vert^2,
\end{equation}
where $\lambda_i$ and $\ket{\psi_i}$ are the eigenvalues and eigenvectors of $\varrho$ respectively, while the Hamiltonian $H$ is the Hermitian generator of the unitary encoding $U = e^{-i H \phi}$, where $\varrho_\phi = U \varrho U^{\dagger}$. Although the POVM does not explicitly depend on $\phi$, in general the optimal POVM, in Eq.~(\ref{qFisher}), may differ for different values of $\phi$. We will discuss further this issue below.

The implication here is that there exists a POVM that will yield the optimal Fisher information, but we do not know much about it -- we shall return to this point later. Let us now go back to the game between Charlie versus Alice and Bob. Again the rule is that Charlie is able to make operations with entangling power, while Alice and Bob do not possess any entangling power. We will shortly show that Charlie's coherent measurement allows better precision in estimating $\phi$ than Alice and Bob's adaptive measurements.

%%%%%%%%%%%%%%%%%%%%%%%%%%%%%%%%%%%%%%%%%%%%%%%%%%%%%%%%%%%%
%%%%%%%%%%%%%%%%%%%%%%%%%%%%%%%%%%%%%%%%%%%%%%%%%%%%%%%%%%%%
\section{Measurements in quantum metrology}
%%%%%%%%%%%%%%%%%%%%%%%%%%%%%%%%%%%%%%%%%%%%%%%%%%%%%%%%%%%%
%%%%%%%%%%%%%%%%%%%%%%%%%%%%%%%%%%%%%%%%%%%%%%%%%%%%%%%%%%%%
To attain the optimal Fisher information requires, in general, an entangling operation involving all parts of the probe. We will refer to such a measurement as a \emph{coherent} measurement. We want to compare the coherent measurement strategy to a measurement strategy where we have no entangling power. Such a strategy then simply involves probabilistic-projective measurements on qubits. We will refer to such a measurement as an \emph{adaptive} measurement. The measurement operators of an adaptive measurement can be decomposed as $\Pi_{a|b} \otimes \Pi_b$. While for a coherent measurement this, in general, cannot be done.

Let the subscripts `co' and `ad' refer to `coherent' and `adaptive' respectively. It is easy to see that $\fisher_{\mbox{co}} \ge \fisher_{\mbox{ad}}$ since the set of coherent measurements contain the set of adaptive measurements. In this article we give examples where $\fisher_{\mbox{co}} > \fisher_{\mbox{ad}}$, which proves that in general entangling power in the readout phase of quantum metrology is necessary for optimal estimation. Also see~\cite{Luo} for a hierarchy of measurements for QFI.

%%%%%%%%%%%%%%%%%%%%%%%%%%%%%%%%%%%%%%%%%%%%%%%%%%%%%%%%%%%%
\subsection{Globally and locally optimal measurements}
%%%%%%%%%%%%%%%%%%%%%%%%%%%%%%%%%%%%%%%%%%%%%%%%%%%%%%%%%%%%
In general Eq.~(\ref{qFisher}) does not tell us much about the optimal POVM. The derivation of Eq.~(\ref{qFisher}) does identify the necessary and sufficient condition for the optimal measurement, namely that 
\begin{equation}\label{opt-povm-conds}
\sqrt{\Pi_x} L_\phi \sqrt{\varrho_\phi} = k_x \sqrt{\Pi_x} \sqrt{\varrho_\phi}
\end{equation}
for constant $k_x \in \mathbb{R}$ and $L_\phi$ is the symmetric logarithmic derivative defined by $\partial_\phi \varrho_\phi = \frac{1}{2}\left(\varrho_\phi L_\phi + L_\phi \varrho_\phi\right)$ ~\cite{Caves}. Although there is no known general solution for the POVM elements $\Pi_x$, a sufficient condition is that they are projectors onto eigenspaces of $L_\phi$. This construction yields measurements that are in general coherent and also local in parameter space (optimal for only certain values of $\phi$).

We do not know whether optimality can be attained with an adaptive measurement nor do we know whether there exists a measurement that is globally optimal, i.e.,for any value of $\phi$. The first issue is inconvenient as we consequently have to search for optimal adaptive measurements on a case-by-case basis, but the second is of purely theoretical interest and of little importance in practice for the following reason. The parameter $\phi$ is estimated from a large number of measurements, $M$, so we can employ an adaptive strategy in which $M'$ ($M' \ll M$) suboptimal measurements are performed to estimate $\phi$, after which we can fix the measurement to be optimal at the approximate parameter value ${\tilde{\phi}}$. It is a well-known result of classical statistics that a maximum likelihood estimator (MLE) for $\phi$ saturates the Cram\'{e}r-Rao bound as $M \rightarrow \infty$, but we can only construct an approximate MLE since we have used a measurement that is optimal at  ${\tilde{\phi}}$ as opposed to $\phi$. Fortuitously, it is nonetheless found to share the same asymptotic properties as the true MLE, provided that $M'$ is sufficiently large, meaning that globally optimal measurements confer no estimation advantage in the limit of large $M$~\cite{Monras, Barndorff-Nielsen}.

In fact, we can reach a useful conclusion regarding the possibility of global optimality: when $\varrho_\phi$ is full-rank (and thus strictly positive) for all $\phi$, we show in \ref{App4} that no globally optimal measurement exists. Moreover, noise perturbs any zero eigenvalues consequently making the result applicable to any practically realisable state. In fact, such a strategy was recently implemented in an experiment reported in \cite{adesso} and it could also be eventually used in a \emph{proof-of-principle} experiment of the ideas introduced here.

%%%%%%%%%%%%%%%%%%%%%%%%%%%%%%%%%%%%%%%%%%%%%%%%%%%%%%%%%%%%
\subsection{Classically correlated quantum states}
%%%%%%%%%%%%%%%%%%%%%%%%%%%%%%%%%%%%%%%%%%%%%%%%%%%%%%%%%%%%
Let us first prove that classically correlated probes do not require any entanglement in the readout. Suppose we use as a probe a bipartite (or multipartite) quantum state that is classically correlated: $\varrho = \sum_{ab} q(a,b) \ket{ab} \bra{ab}$ which is mapped to $\varrho_\phi = \sum_{a_\phi b_\phi} q(a,b) \ket{a_\phi b_\phi} \bra{a_\phi b_\phi}$ due to the encoding, where $\ket{j_\phi} = U \ket{j}$ for $j=a,b$. Now, note that this state is invariant under local projective measurements
\begin{equation}\label{Classical-Cond}
\varrho_\phi = \sum_{a_\phi b_\phi} \ket{a_\phi b_\phi} \bra{a_\phi b_\phi} \varrho_\phi \ket{a_\phi b_\phi} \bra{a_\phi b_\phi},
\end{equation}
which is the definition of a classically correlated state~\cite{Discord3, Discord4}. However, there exists a measurement $\{\Pi_x\}$ that yields the outcome probabilities $p_{\phi}(x) = \tr[\Pi_x \varrho_\phi]$ leading to optimal QFI. Using Eq.~(\ref{Classical-Cond}) and the cyclic invariance of the trace we get
\begin{equation}\label{clprob}
p_{\phi}(x) = \sum_{a_\phi b_\phi} \braket{a_\phi b_\phi| \Pi_x | a_\phi b_\phi} \tr\left[\ket{a_\phi b_\phi}  \bra{a_\phi b_\phi} \varrho_\phi \right].
\end{equation}
It is then straightforward to write the POVM elements as 
\begin{equation}\label{clpovm}
\Pi_{x} = \sum_{a_\phi b_\phi} \braket{a_\phi b_\phi| \Pi_x | a_\phi b_\phi} \ket{a_\phi b_\phi} \bra{a_\phi b_\phi},
\end{equation}
which is diagonal in a product basis.

Next, we assume that Alice and Bob have full knowledge of the elements $\Pi_x$. In order to implement $\Pi_x$ they would need to implement entangling operations. However, instead they may start with an estimate $\tilde{\phi}$ of the phase and apply the following POVM
\begin{equation}\label{clpovmest}
\tilde\Pi_{x} = \sum_{a_{\tilde\phi} b_{\tilde\phi}} \braket{a_{\tilde\phi} b_{\tilde\phi}| \Pi_x | a_{\tilde\phi} b_{\tilde\phi}} \ket{a_{\tilde\phi} b_{\tilde\phi}} \bra{a_{\tilde\phi} b_{\tilde\phi}}.
\end{equation}
Since they know the POVM element they can easily compute the probability distribution $\braket{a_{\tilde\phi} b_{\tilde\phi}| \Pi_x | a_{\tilde\phi} b_{\tilde\phi}}$. Then they simply make measurements in basis $\ket{a_{\tilde\phi} b_{\tilde\phi}}$, which is local. After each round they can update the value of $\tilde\phi$ so that it is closer to the actual value of $\phi$. In this way they can adaptively approach the POVM in Eq.~(\ref{clpovm}) using local measurements and classical communication.

This means that the such a measurement can be implemented without the aid of an entangling operation. Our main point here is that if $\varrho_\phi$ is not classical then it may not be possible to reproduce the optimal distribution $p_{\phi}(x)$ via local measurements on $A$ and $B$. On the other hand, for classical probes states the POVM elements, as given in Eq.~(\ref{clpovm}), can be implemented locally. A word of caution is necessary here. In general implementing a separable POVM may require entanglement \cite{NwoE}. However, in the construction above, the basis vectors of the POVM are locally distinguishable, i.e., $\braket{a|a'} = \delta_{aa'}$ and $\braket{b|b'} = \delta_{bb'}$. Therefore it is possible to implement this POVM via LOCC. 

Finally, note that this result is fully independent of the formula for QFI in Eq.~(\ref{qFisher}). Moreover, it might seem that in order for Alice and Bob to implement the POVM given above, they need to know the value of $\phi$. It is important to keep in mind that the value of $\phi$ is not totally unknown, and it is the precision in the value of $\phi$ that we desire. Therefore it is reasonable to assume that Alice and Bob know the neighbourhood in which $\phi$ lies and implement $\Pi_{a,b,x}^\phi$ with increasing accuracy.

%%%%%%%%%%%%%%%%%%%%%%%%%%%%%%%%%%%%%%%%%%%%%%%%%%%%%%%%%%%%
\subsection{Quantum correlated N-GHZ and $N00N$ states}
%%%%%%%%%%%%%%%%%%%%%%%%%%%%%%%%%%%%%%%%%%%%%%%%%%%%%%%%%%%%
Before we give an example exhibiting a gap between the coherent and adaptive Fisher information, let us discuss the important case of $N00N$ states. A $N00N$ state is an optical state of $N$ photons in superposition with the vacuum in two arms of an interferometer: $(\ket{N0}+\ket{0N})/\sqrt{2}$. For linear encoding, $N00N$ states saturate the ultimate bound for Fisher information, the Heisenberg limit $N^2$~\cite{Maccone}. 

Here we will work with $N-$qubit GHZ (NGHZ) states instead of $N00N$ states, but in principal the two are equivalent~\cite{NOON, NOON2, NOON3, NOON4}. An NGHZ state before the encoding is where all qubits are in the state $\ket{0}$ in superposition with all qubits in state $\ket{1}$. It is written as $\ket{G^N} = ( \ket{0}^{\otimes N} + \ket{1}^{\otimes N} ) /\sqrt{2}$ which is transformed to $\ket{G_\phi^N} = (\ket{0}^{\otimes N} + e^{i N \phi} \ket{1}^{\otimes N})/\sqrt{2}$ after the encoding. That is, the phase is encoded by $U^{\otimes N}$ with $U = e^{i \phi \ket{1}\bra{1}}$. Here, $\ket{0}$ and $\ket{1}$ are the eigenstates of the Pauli matrix $\sigma_{z}$.

The coherent readout of the latter state requires a series of {\sc C-not} gates between the first qubit and the remaining $N-1$ qubits, resulting in the state $(1/\sqrt{2})(\ket{0} + e^{i N \phi} \ket{1}) \otimes \ket{0}^{\otimes N-1}$. The first qubit can now be measured in the $\ket{\pm} = \left(\ket{0} \pm \ket{1}\right)/\sqrt{2}$ basis and the desired Fisher information is achieved. However, the same result is achieved adaptively by measuring each qubit in the $\ket{\pm}$ basis. After $N-1$ qubits are measured the state of the remaining qubit is $(\ket{0} + (-1)^k e^{i N \phi} \ket{1})/\sqrt{2}$, where $k$ is the number of times the outcome $\ket{-}$ was observed. If $k$ is odd we apply a $\sigma_z$ operation, but no correction is required otherwise.

Strangely, this example shows that coherent processing is unnecessary for readout in metrology with NGHZ states. It seems that for \emph{pure} probes the adaptive strategy could be equivalent to the coherent strategy, as hinted at in~\cite{Maccone}. We were unable to prove this, nor able to find a counterexample. However, in a real experiment one never has a pure state, so we will now introduce the addition of noise to the problem.

%%%%%%%%%%%%%%%%%%%%%%%%%%%%%%%%%%%%%%%%%%%%%%%%%%%%%%%%%%%%
\subsection{Werner state}
%%%%%%%%%%%%%%%%%%%%%%%%%%%%%%%%%%%%%%%%%%%%%%%%%%%%%%%%%%%%
Let us now choose the probe as the Werner state $\mathcal{W}= (1-\eta) \frac{\mathbb{I}}{4}  + \eta \ket{\mathcal{B}_{00}} \bra{\mathcal{B}_{00}} $, where $\ket{\mathcal{B}_{00}} = \left(\ket{00} + \ket{11} \right)/\sqrt{2}$. The parameter $\eta \in [0,1]$ is the strength of the signal, while $1-\eta$ indicates the amount of white noise present in the probe state. Using Eq.~(\ref{qFisher}) we can compute the optimal Fisher information that Charlie can attain.
\begin{equation}
\fisher_{\mbox{co}}(\mathcal{W}) = \frac{8 \eta^2}{1+\eta}.
\end{equation}

Now, we need to compare this result to the adaptive strategy. Bob's local Fisher information vanishes because his local state is maximally mixed: He gets completely random bits for \emph{any} POVM he implements. Furthermore, we can fine-grain the POVM to a projector onto the state $\ket{\beta^*} = b^*_0 \ket{0} + b^*_1 \ket{1}$. The corresponding conditional states of Alice are
$\mathcal{W}^{A|\beta}_\phi = (1-\eta) \frac{\mathbb{I}}{2} + \eta \ket{\beta_\phi} \bra{\beta_\phi}$, where $\ket{\beta_\phi} = b_0 \ket{0} + e^{i 2 \phi} b_1 \ket{1}$.

The conditional state of Alice is as if Alice had prepared a state $\mathcal{W}^{A|\beta} = (1-\eta) \frac{\mathbb{I}}{2} + \eta \ket{\beta} \bra{\beta}$ and sent it through an interferometer with the phase generated by the Hamiltonian $2 \ket{1}\bra{1}$. Therefore we can simply compute the QFI using Eq.~(\ref{qFisher}).

We know that the QFI will be maximised for $b_0 = b_1 = 1/\sqrt{2}$, therefore we can conclude that Bob will make measurements $\ket{\pm}$. A $\sigma_z$ operation is applied to Alice's state when Bob's outcome is minus to make the two conditional states the same. The adaptive Fisher information is
\begin{equation}
\fisher_{\mbox{ad}}(\mathcal{W}) = 4 \eta^2.
\end{equation}

Note that there is a finite difference between the coherent readout and adaptive readout. The difference vanishes only when $\eta = 0 \, \mbox{or} \,1$. Therefore, in this example, Charlie's ability to perform coherent interactions plays a non-trivial role in phase estimation in the presence of noise. This is our central result, but we will discuss its consequence after we generalise to the case of $N$ qubits.
 
%%%%%%%%%%%%%%%%%%%%%%%%%%%%%%%%%%%%%%%%%%%%%%%%%%%%%%%%%%%%
\subsection{Multipartite states}
%%%%%%%%%%%%%%%%%%%%%%%%%%%%%%%%%%%%%%%%%%%%%%%%%%%%%%%%%%%%
We can represent a multipartite probe with white noise as $\mathcal{W}^N = (1-\eta) \frac{\mathbb{I}}{2^{N}} + \eta \ket{G^N} \bra{G^N}$ with $\ket{G^N)} = (\ket{0}^{\otimes N} + \ket{1}^{\otimes N} )/\sqrt{2}$. The phase shift is now introduced by the unitary $U^{\otimes N}$ with $U = e^{i\phi \ket{1}\bra{1}}$. Charlie's Fisher information is computed using Eq.~(\ref{qFisher}) (see~\ref{App2} for the details):
\begin{equation}\label{coherent}
\fisher_{\mbox{co}}(\mathcal{W}^N) = \frac{2^N} {2^N \eta + 2(1-\eta)} N^2 \eta^2.
\end{equation}

For the adaptive strategy, once again, the local Fisher information for any party is null since the local states are maximally mixed. The sequence of adaptive measurements on the space spanned by the $N$-parties is equivalent to Alice applying $N$ times the phase on her qubit, resulting in the conditional Fisher information:
\begin{equation}\label{adaptive}
\fisher_{\mbox{ad}} (\mathcal{W}^N)=  N^2 \eta^2
\end{equation}
which is greater than zero for a noisy probe ($\eta \neq 0,1$). The derivation of Eqs.~(\ref{coherent}) and (\ref{adaptive}) via a direct calculation of the QFI can be found in~\ref{App2}. On the other hand in~\ref{App3}, we provide an example where the optimal POVM for both coherent and adaptive strategies is obtained for a given value of $\phi$ considering a phase estimation protocol. The POVM approach presented in~\ref{App3} is shown to yield the same result as the derivation based on QFI.

%%%%%%%%%%%%%%%%%%%%%%%%%%%%%%%%%%%%%%%%%%%%%%%%%%%%%%%%%%%%
%%%%%%%%%%%%%%%%%%%%%%%%%%%%%%%%%%%%%%%%%%%%%%%%%%%%%%%%%%%%
\section{Analysis and conclusions}
%%%%%%%%%%%%%%%%%%%%%%%%%%%%%%%%%%%%%%%%%%%%%%%%%%%%%%%%%%%%
%%%%%%%%%%%%%%%%%%%%%%%%%%%%%%%%%%%%%%%%%%%%%%%%%%%%%%%%%%%%

Above, the coherent strategy offers a quadratic enhancement over the adaptive strategy in $\eta$. The ratio of the two Fisher information amounts to the number of times the adaptive strategy has to be repeated to match the precision attained by the coherent strategy. Even for a handful of qubits we have $1-\eta \ll 2^{N-1}$ and
\begin{equation}
\fisher_{\mbox{ad}}(\mathcal{W}^N) 
= \left(\eta + \frac{1-\eta} {2^{N-1}} \right) \fisher_{\mbox{co}}(\mathcal{W}^N) 
\approx \eta \, \fisher_{\mbox{co}}(\mathcal{W}^N).
\end{equation}
For highly mixed states with a very small value of $\eta$, the adaptive readout performs extremely poorly compared to the coherent readout. This has huge implications for magnetic field sensing in the nuclear magnetic resonance setup~\cite{Jones, Simmons}, where $\eta \approx 10^{-5}$ and implies more than three-hundred times better precision in $\phi$ due to coherent operations. 

It is important to note that at small values of $\eta$ there is no entanglement in the probe state. Therefore the effect we uncover here has little to do with entanglement and more to do with operations that have the potential to generate entanglement.

Interestingly, the gap in the Fisher information is reminiscent of the gap in the Holevo quantity between coherent decoding versus adaptive decoding, which is shown to be quantum discord in~\cite{Gu, weedbrook}. There are several characteristic traits of quantum mechanics that distinguish it from the classical theory: besides non-classical correlations like entanglement and discord, the possibility of performing coherent interactions between different partitions of the probe does not have a classical analogue~\cite{ModiGu, Filgueiras}.

In this article we showed that, in general, coherent readout leads to better precision over adaptive readout in quantum parameter estimation. We also showed that the two readouts are equivalent for classical probe states. Finally, the noise in some quantum correlated probes can be quadratically suppressed with the use of quantum coherent operations, leading to better precision for parameter estimation. In this manner we have highlighted the importance of coherent measurements in quantum metrology in the presence of quantum correlations.

\ack

LC, KMi, and RS are grateful for the financial support from CNPq, CAPES, and INCT-IQ. KMi and RS are grateful for the financial support from FAPESP. DR gratefully acknowledges the hospitality of the University of Oxford, and the Institute of Physics and Nuffield Foundation for financial support. FAP thanks the Leverhulme Trust for financial support. The John Templeton Foundation, the National Research Foundation, and the Ministry of Education of Singapore supported KMo during the completion of this work. KMo thanks UFABC for their hospitality.

\appendix
%%%%%%%%%%%%%%%%%%%%%%%%%%%%%%%%%%%%%%%%%%%%%%%%%%%%%%%%%%%%
%%%%%%%%%%%%%%%%%%%%%%%%%%%%%%%%%%%%%%%%%%%%%%%%%%%%%%%%%%%%
\section{Classical-additivity of Fisher information}\label{App1}
%%%%%%%%%%%%%%%%%%%%%%%%%%%%%%%%%%%%%%%%%%%%%%%%%%%%%%%%%%%%
%%%%%%%%%%%%%%%%%%%%%%%%%%%%%%%%%%%%%%%%%%%%%%%%%%%%%%%%%%%%
In this section we will derive Eq.~(5) of the main text and its generalisation to a multipartite case will be obtained. From conditional probability, we can write the bipartite probability distribution as $p_{\phi}(a,b) = p_{\phi}(a|b) p_{\phi}(b)$ and its associated Fisher information reads
\begin{eqnarray}
\fisher(A,B) &=& \sum_{a,b} p_{\phi}(a,b) \left[\partial_\phi \ln(p_{\phi}(a,b)) \right]^2 
=\sum_{a,b} p_{\phi}(a,b) \left[ \frac{\partial_\phi p_{\phi}(a,b)}{p_{\phi}(a,b)}  \right]^2 \nonumber\\
&=& \sum_{a,b} p_{\phi}(a|b)p_{\phi}(b) \left[ \frac{\partial_\phi [p_{\phi}(a|b)p_{\phi}(b)]}{p_{\phi}(a|b)p_{\phi}(b)} \right]^2 \nonumber\\
&=& \sum_{a,b} p_{\phi}(a|b)p_{\phi}(b) \left( \frac{\partial_\phi p_{\phi}(a|b)}{p_{\phi}(a|b)} + \frac{\partial_\phi p_{\phi}(b)}{p_{\phi}(b)}  \right)^2 \nonumber\\
&=& \sum_{a} p_{\phi}(a|b) \sum_{b} \frac{\left[\partial_\phi p_{\phi}(b) \right]^2}{p_{\phi}(b)} + \sum_{b} p_{\phi}(b) \sum_{a} \frac{\left[\partial_\phi p_{\phi}(a|b) \right]^2}{p_{\phi}(a|b)} \nonumber\\
&&+ 2 \sum_{b} {\partial_\phi}p_{\phi}(b) \sum_{a} {\partial_\phi} p_{\phi}(a|b).
\end{eqnarray}
Now note that $\sum_{a} p_{\phi}(a|b) = p_{\phi}(b)/p_{\phi}(b) =1$ and therefore $\sum_a {\partial_\phi} p_{\phi}(a|b) = {\partial_\phi} \sum_a p_{\phi}(a|b) = 0$, we obtain
\begin{eqnarray}
\fisher(A,B) &=& \sum_{a} p_{\phi}(a|b) \sum_{b} \frac{\left[{\partial_\phi} p_{\phi}(b) \right]^2}{p_{\phi}(b)}  + \sum_{b} p_{\phi}(b) \sum_{a} \frac{\left[{\partial_\phi} p_{\phi}(a|b) \right]^2}{p_{\phi}(a|b)}  \nonumber\\
&=& \fisher(B) + \sum_b p_{\phi}(b) \fisher(A|B=b) = \fisher(B) + \fisher(A|B),
\end{eqnarray}
which is the Eq.~(5) of the main text.  

%%%%%%%%%%%%%%%%%%%%%%%%%%%%%%%%%%%%%%%%%%%%%%%%%%%%%%%%%%%%
%%%%%%%%%%%%%%%%%%%%%%%%%%%%%%%%%%%%%%%%%%%%%%%%%%%%%%%%%%%%
\subsection{Multipartite case}\label{mult}
%%%%%%%%%%%%%%%%%%%%%%%%%%%%%%%%%%%%%%%%%%%%%%%%%%%%%%%%%%%%
%%%%%%%%%%%%%%%%%%%%%%%%%%%%%%%%%%%%%%%%%%%%%%%%%%%%%%%%%%%%
Now, let us consider the three partition case ($a,b$ and $c$) with joint probability distribution $p_{\phi}(a,b,c)$. In this case two successive applications of the definition of conditional probability result in $p_{\phi}(a,b,c) = p_{\phi}(a|bc) p_{\phi}(b|c) p_{\phi}(c)$. In this fashion, we can write
\begin{eqnarray*}
\fisher(A,B,C) &=& \sum_{abc} p_{\phi}(a,b,c) \left[ \frac{{\partial_\phi} p_{\phi}(a,b,c)}{p_{\phi}(a,b,c)} \right]^2 \nonumber\\
&=& \sum_{abc} p_{\phi}(a|bc)p_{\phi}(b|c)p_{\phi}(c) \left[ \frac{{\partial_\phi} [p_{\phi}(a|bc) p_{\phi}(b|c)p_{\phi}(c)]} {p_{\phi}(a|bc)p_{\phi}(b|c)p_{\phi}(c)} \right]^2,
\end{eqnarray*}
which after some algebra turns out to be
\begin{eqnarray}
\fisher(A,B,C) &=& \sum_c \frac{\left[ {\partial_\phi} p_{\phi}(c) \right]^2}{p_{\phi}(c)} + \sum_c p_{\phi}(c) \sum_b \frac{\left[{\partial_\phi}p_{\phi}(b|c) \right]^2}{p_{\phi}(b|c)}
\nonumber\\&& + \sum_{bc} p_{\phi}(bc) \sum_a \frac{\left( {\partial_\phi} p_{\phi}(a|bc) \right)^2}{p_{\phi}(a|bc)} \nonumber\\
&=& \fisher(C) + \fisher(B|C) + \fisher(A|BC). \nonumber
\end{eqnarray}

By successive applications of the definition of conditional probability, it is easy to show that for a multipartite probe (with $N$ partitions $\{X_{1},\dots,X_{N}\}$), we can write a fancy generalisation as
\begin{equation}\label{manyrounds}
\fisher(X_1 \dots X_N) = \fisher(X_N) + \sum_{k=1}^{N-1} \fisher(X_k | X_{k+1} \dots X_N). 
\end{equation}

%%%%%%%%%%%%%%%%%%%%%%%%%%%%%%%%%%%%%%%%%%%%%%%%%%%%%%%%%%%%
%%%%%%%%%%%%%%%%%%%%%%%%%%%%%%%%%%%%%%%%%%%%%%%%%%%%%%%%%%%%
\section{Fisher information for $\mathcal{W}^N$}\label{App2}
%%%%%%%%%%%%%%%%%%%%%%%%%%%%%%%%%%%%%%%%%%%%%%%%%%%%%%%%%%%%
%%%%%%%%%%%%%%%%%%%%%%%%%%%%%%%%%%%%%%%%%%%%%%%%%%%%%%%%%%%%

%%%%%%%%%%%%%%%%%%%%%%%%%%%%%%%%%%%%%%%%%%%%%%%%%%%%%%%%%%%%
\subsection{Coherent Fisher information}
%%%%%%%%%%%%%%%%%%%%%%%%%%%%%%%%%%%%%%%%%%%%%%%%%%%%%%%%%%%%

Here we wish to compute the quantum Fisher information (QFI) with coherent readout for the following state 
\begin{equation}
\mathcal{W}^{N} = (1-\eta) \frac{\mathbb{I}}{2^N} + \eta \ket{G^N} \bra{G^N},
\label{Wnstate}
\end{equation}
where
$ \ket{G^N} = (\ket{0}^{\otimes N} + \ket{1}^{\otimes N}) / \sqrt{2}.$ To do this we need the eigenvectors and eigenvalues of the state above. Eigenvector $\ket{G^N}$ comes with eigenvalue $\eta + (1-\eta)/2^N$ and all other $2^N -1$ eigenvectors come with eigenvalue $(1-\eta)/2^N$. Next note that the Hamiltonian that encodes the parameter to be estimated is $H_N = \bigoplus_i H_i$ with $H_i = \ket{1}\bra{1}$.

Now we simply utilise Eq.~(6) in main text and compute QFI. We begin by noting that the eigenstates with the same eigenvalues do not contribute and the action of the Hamiltonian on $\ket{G^N}$ is $(1/\sqrt{2} \ket{1}^{\otimes N})$. Therefore the only other eigenvector that matters is $\ket{\bar{G}^N}  = (\ket{0}^{\otimes N} - \ket{1}^{\otimes N}) / \sqrt{2}$. This is because all other eigenstates of $\varrho_N$ are orthonormal to the $\ket{1}^{\otimes N}$ term. The QFI is then:
\begin{eqnarray}
\fisher_{\mbox{co}}(\mathcal{W}^{N}) &=& 4 \frac{\left(\eta + (1-\eta)/2^N - (1-\eta)/2^N \right)^2}{\eta + (1-\eta)/2^N + (1-\eta)/2^N} \left|\braket{\bar{G}^N| H_N | G^N}\right|^2 \\
&=& \frac{\eta^2}{\eta + 2(1-\eta)/2^N }
\left|\left(\bra{0}^{\otimes N} - \bra{1}^{\otimes N}\right) N \ket{1}^{\otimes N}\right|^2\\
&=& \frac{2^N }{2^N\eta + 2(1-\eta)}  N^2 \eta^2.
\end{eqnarray}

%%%%%%%%%%%%%%%%%%%%%%%%%%%%%%%%%%%%%%%%%%%%%%%%%%%%%%%%%%%%
\subsection{Adaptive Fisher information}
%%%%%%%%%%%%%%%%%%%%%%%%%%%%%%%%%%%%%%%%%%%%%%%%%%%%%%%%%%%%

Now let us compute the QFI for adaptive readout. For $N$ qubits the adaptive strategy reduces to making (probabilistic) projective measurements. We begin by considering the measurement by the first qubit. Since the local state of this qubit is maximally mixed, the local Fisher information vanishes. Therefore, to maximise the Fisher information for adaptive readout, the Fisher information of the remaining $N-1$ qubits must be maximised.

After the parameter has been encoded the state of $N$ qubits is
\begin{eqnarray}
\mathcal{W}^{N}_\phi &=& (1-\eta) \frac{\mathbb{I}}{2^{N}} + \eta \frac{1}{2} \left(\ket{0}^{\otimes N} + e^{i N \phi} \ket{1}^{\otimes N} \right) \nonumber\\
&& \times \left(\bra{0}^{\otimes N} + e^{-i N \phi} \bra{1}^{\otimes N} \right).\label{nqubitstate}
\end{eqnarray}
Let us imagine that the last party has a measurement outcome along some direction $\bra{m} = m_0 \bra{0} + m_1 \bra{1}$, with $|m_0|^2 + |m_1|^2 = 1$. The corresponding conditional state of the remaining qubits is
\begin{eqnarray}
\mathcal{W}^{N-1}_\phi &=& (1-\eta) \frac{\mathbb{I}}{2^{N-1}} + \eta \left(m_0 \ket{0}^{\otimes N-1} + m_1 e^{i N \phi} \ket{1}^{\otimes N-1} \right) \nonumber\\
&& \times\left(m^*_0 \bra{0}^{\otimes N-1} + m^*_1 e^{-i N \phi} \bra{1}^{\otimes N-1} \right)
\end{eqnarray}
We want to find values of $m_0$ and $m_1$ that maximise the Fisher information of the last state. We note that the last state is exactly as if we had prepared the state of $N-1$
\begin{eqnarray}
\mathcal{W}^{N-1} &=& (1-\eta) \frac{\mathbb{I}}{2^{N-1}} + \eta \left(m_0 \ket{0}^{\otimes N-1} + m_1  \ket{1}^{\otimes N-1} \right) \nonumber\\ && \times
 \left(m^*_0 \bra{0}^{\otimes N-1} + m^*_1 \bra{1}^{\otimes N-1} \right)
\end{eqnarray}
and sent it through a process generated by the Hamiltonian $\frac{N}{N-1} \bigoplus_i \ket{1}\bra{1}$. The coherent QFI of the last state can be computed to be
\begin{equation}
\fisher_{\mbox{co}}(\mathcal{W}^{N-1})
= 4 \frac{2^{N-1} }{2^{N-1}-(2^{N-1}-2)(1-\eta)} N^2 |m_0 m_1|^2 \eta^2,
\end{equation}
which is maximum for $m_0 = m_1 = 1/\sqrt{2}$. Therefore we can conclude that the first measurement is in the $\ket{\pm}$ basis. The resultant conditional state of the $N-1$ qubits is
\begin{eqnarray}
\mathcal{W}^{N-1}_\phi &=& (1-\eta) \frac{\mathbb{I}}{2^{N-1}} + \eta \frac{1}{2} \left( \ket{0}^{\otimes N-1} \pm e^{i N \phi} \ket{1}^{\otimes N-1} \right) \nonumber\\&& \times
\left(\bra{0}^{\otimes N-1} \pm e^{-i N \phi} \bra{1}^{\otimes N-1} \right).
\end{eqnarray}
But this state looks exactly like the $N$ qubit state we started with in Eq.~(\ref{nqubitstate}). Therefore by carrying out the exact same analysis we find that the next measurement also has to be in the $\ket{\pm}$ basis.

After all but the last party has measured the final qubit has the state
\begin{equation}
\mathcal{W}^{1}_\phi = (1-\eta) \frac{\mathbb{I}}{2} + \eta \frac{1}{2} \left( \ket{0} \pm e^{i N \phi} \ket{1} \right) \left(\bra{0} \pm e^{-i N \phi} \bra{1}\right).
\end{equation}
Let us denote the number of parties that observe $\ket{-}$ with $k$. If $k$ is odd then a $\sigma_z$ is applied to change the minus sign in the superposition. This state is exactly as if we had prepared the state
\begin{equation}
\mathcal{W}^{1}_\phi = (1-\eta) \frac{\mathbb{I}}{2} + \eta \frac{1}{2} \left( \ket{0} + \ket{1} \right) \left(\bra{0} + \bra{1}\right)
\end{equation}
and sent it through a process generated by Hamiltonian $N \ket{1}\bra{1}$. The Fisher information for this Hamiltonian and the last state is the Fisher information for adaptive readout:
\begin{equation}
\fisher_{\mbox{ad}}(\mathcal{W}^{N}) = \frac{\eta^2}{\eta + (1-\eta)} N^2 = \eta^2 N^2.
\end{equation}

%%%%%%%%%%%%%%%%%%%%%%%%%%%%%%%%%%%%%%%%%%%%%%%%%%%%%%%%%%%%
%%%%%%%%%%%%%%%%%%%%%%%%%%%%%%%%%%%%%%%%%%%%%%%%%%%%%%%%%%%%
\section{Explicit example of the optimal POVM for $\mathcal{W}^N$}\label{App3}
%%%%%%%%%%%%%%%%%%%%%%%%%%%%%%%%%%%%%%%%%%%%%%%%%%%%%%%%%%%%
%%%%%%%%%%%%%%%%%%%%%%%%%%%%%%%%%%%%%%%%%%%%%%%%%%%%%%%%%%%%

In this section we will illustrate how to build the optimal POVM through a specific phase estimation example for the probe state $\mathcal{W}^N$ in Eq.~(\ref{Wnstate}) and the generator $H_N$ defined previously. Here, we will employ the general Fisher information expression, as given in Eq.~(\ref{jointFisher}), considering a probability distribution obtained by a POVM instead of using the QFI. We will also show that in this example the measurement is optimal for a given value of the parameter $\phi$ resulting in the same gap between the global and local strategies obtained in the previous appendix. Let us start from the encoded probe state given by
\begin{equation}
\mathcal{W}^{N}_{\phi} = U^{\otimes N} \mathcal{W}^{N} U^{\dagger \otimes N}  = (1-\eta)\frac{\mathbb{I}}{2^{N}} + \eta \ket{G_\phi^N} \bra{G_\phi^N}, \label{rhophi}
\end{equation} 
with $\ket{G_\phi^N} = \left(\ket{0}^{\otimes N}+e^{iN\phi}\ket{1}^{\otimes N} \right)/\sqrt{2}$.  

%%%%%%%%%%%%%%%%%%%%%%%%%%%%%%%%%%%%%%%%%%%%%%%%%%%%%%%%%%%%
\subsection{Global coherent strategy}
%%%%%%%%%%%%%%%%%%%%%%%%%%%%%%%%%%%%%%%%%%%%%%%%%%%%%%%%%%%%

%%%%%%%%%%%%%%%%%%%%%%%%%%%%%%%%%%%%%%%%%%%%%%%%%%%%%%%%%%%%
\begin{figure}
\begin{center}
\includegraphics[scale=.45]{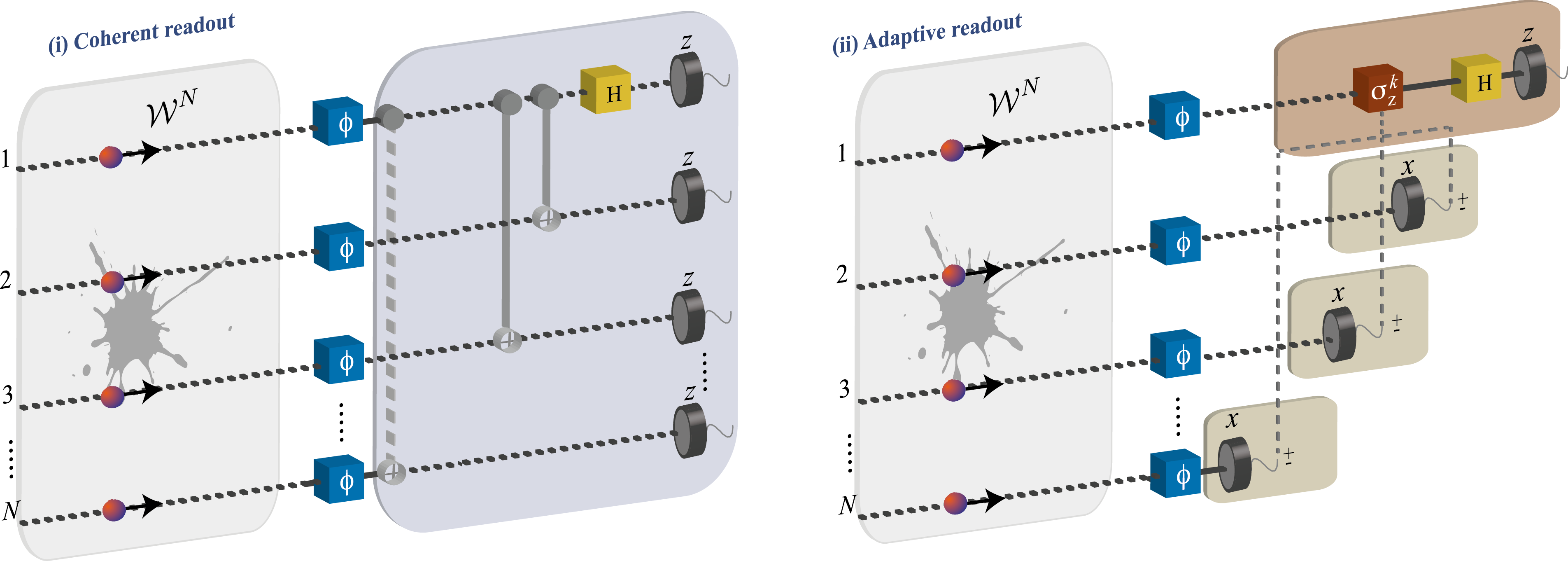}
\end{center}
\caption{\label{StratFigMP}Two readout strategies for parameter estimation for the multipartite case with probe state $\mathcal{W}^N$. In (i) the coherent readout process consists of applying a series of \textsc{C-not} gates, which are entangling operations, followed by measuring all partitions. In (ii) each party measures their partition in the $\ket{\pm}$ basis and communicates the outcome to the first party. The first party applies a $\sigma_z$ if the number of $\ket{-}$ outcomes was odd. Finally, she makes a measurement on her qubit. The key ingredient for coherent readout is the ability to perform an entangling unitary transformation, which is the \textsc{C-not} gate here.}
\end{figure}
%%%%%%%%%%%%%%%%%%%%%%%%%%%%%%%%%%%%%%%%%%%%%%%%%%%%%%%%%%%%

As in any practical scenario we will restrict our analysis to some class of POVMs. Note that in a given experimental setup the kind of operations that we can perform without adding ancillary systems is limited by the nature of the setup. We will start from the optimal POVM for the global coherent strategy in the the noiseless case (outlined in Fig.~\ref{StratFigMP}(i)). We will further see that this choice of POVM is optimal for a specific value of $\phi$. Indeed, this POVM is the set of projectors on the Bell basis state (note that $\mathcal{W}^{N}$ is a \textit{Bell diagonal state}). Such a readout can be constructed by means of a sequence of coherent interactions (C-{\sc not} gates) as indicated  in Fig.~\ref{StratFigMP}(i). We can compactly represent the $N$ C-{\sc not} gates in the readout strategy as
\begin{equation}
{\rm{C}}^N_{\rm NOT} = \ket{0}\bra{0} \otimes \mathbb{I}_2^{\otimes N-1} + \ket{1}\bra{1} \otimes \sigma_x^{\otimes N-1},
\end{equation}
so that 
\begin{eqnarray}
\label{rhoCnot}
{\rm{C}}_{\rm NOT}^N \mathcal{W}^{N}_{\phi} {\rm{C}}_{\rm NOT}^N %
&=& \left[ \eta \ket{G_\phi}\bra{G_\phi} + (1-\eta) \frac{\mathbb{I}_{2}}{2^N} \right] \otimes\ket{0}\bra{0}^{\otimes N-1} \nonumber\\
&&+ (1-\eta)\frac{\mathbb{I}_{2}}{2^N}\otimes \left( \mathbb{I}_2^{\otimes N-1} - \ket{0}\bra{0}^{\otimes N-1} \right),
\end{eqnarray}
with $\ket{G_\phi} = \left(\ket{0}+e^{iN\phi}\ket{1} \right)/\sqrt{2}$. The global POVM set can be written as {$\{\Pi^G_k = {\rm{C}}_{\rm NOT}^N H \ket{k_{({\rm bin})}} \bra{k_{({\rm bin})}} H {\rm C}_{\rm NOT}^N\}$}, where $k = 0,1,2,\dots,(2^N -1)$ and $k_{({\rm bin})}$ is the binary representation of $k$ with $N$-bits. In this readout scheme the obtained probability distribution reads
\begin{equation}
p_\phi (k) = {\rm Tr~} \big( \mathcal{W}^{N}_{\phi} \Pi^G_k \big).
\end{equation}
Since $\frac{\partial}{\partial\phi} p_\phi(k) = 0$ for $k \neq 0, 2^{N-1}$, the global Fisher information, in this case, is given by 
\begin{equation}
\mathcal{F}_{\mbox{co}} = \sum_k \frac{[\partial_\phi p_\phi(k)]^2}{p_\phi(k)} = \frac{[\partial_\phi p_\phi(0)]^2}{p_\phi(0)} + \frac{[\partial_\phi p_\phi(2^{N-1})]^2}{p_\phi(2^{N-1})},
\end{equation}
where $p_\phi(0) = \eta\cos^2\left( N \frac{\phi}{2} \right) + \frac{1-\eta}{2^N}$ and $p_\phi(2^{N-1}) = \eta\sin^2\left( N \frac{\phi}{2} \right) + \frac{1-\eta}{2^N}$. From which it follows that
\begin{equation}
\max_{\{\phi\}}\mathcal{F}_{\mbox{co}} = \frac{2^N}{2^N \eta + 2(1-\eta)} N^2 \eta^2.
\label{POVMcoher}
\end{equation}

%%%%%%%%%%%%%%%%%%%%%%%%%%%%%%%%%%%%%%%%%%%%%%%%%%%%%%%%%%%%
\subsection{Adaptive strategy}
%%%%%%%%%%%%%%%%%%%%%%%%%%%%%%%%%%%%%%%%%%%%%%%%%%%%%%%%%%%%

In the adaptive strategy, the set of local POVMs corresponding to the circuit displayed in Fig.~\ref{StratFigMP}(ii) is $\left\{ \Pi^{(x_1)}_{1|2 \dots N} \otimes \left( \bigotimes_{k=2}^N \Pi^{(x_k)}_{k} \right) \right\}$. The measurement on the qubits $2,3,\dots, N$ is a projective measurement performed in the $\sigma_x$ basis and $\Pi^{(x_k)}_{k} = \ket{x_k}\bra{x_k}$ for $k\geq 2$, where $x_k=+,-$. This is an optimal POVM for the noiseless scenario. A convenient form for these projectors is given by $\Pi^{(x_k)}_{k} = \big( \mathbb{I}_2 + (-1)^{f(x_k)} \sigma_x \big) / 2$ with $f(x_k)=0,1$. Therefore, the POVM set for the adaptive strategy can be written as 
\begin{equation}
\label{adptpovm}
\left\{ \Pi_{\mbox{ad}} = \sigma_z^\mu H \ket{x_1}\bra{x_1} H \sigma_z^\mu \otimes  \left( \bigotimes_{k=2}^N \Pi^{(x_k)}_{k}  \right) \right\},
\end{equation}
where $x_1=0,1$ and \hbox{$\sum_{k=2}^N f(x_k) \equiv \mu \pmod{2}$} is the parity of the outcomes $\{x_2,\dots,x_N\}$.

Let us inspect the action of these measurements on the state $\mathcal{W}^{N}_{\phi}$ given in Eq.~\ref{rhophi}. Considering the operation $\Pi_N^{(x_N)}$ on the $N$th qubit, we have

\begin{equation}
\Pi_N^{(x_N)} \mathcal{W}^{N}_{\phi} \Pi_N^{(x_N)}%
= \left[ \eta \ket{G_\phi^{N-1}}\bra{G_\phi^{N-1}} + (1-\eta)\frac{\mathbb{I}_2^{\otimes N-1}}{2^{N-1}} \right] \otimes \frac{\Pi_N^{(x_N)}}{2},
\end{equation}
where $\ket{G_\phi^{N-1}} = \left(\ket{0}^{\otimes N-1} +(-1)^{f(x_N)} e^{iN\phi}\ket{1}_{\otimes N-1} \right)/\sqrt{2}$ and $x_N$ is the outcome of the measurement on the $N$th qubit. Now, applying the operations $\bigotimes_{k=2}^{N-1} \Pi^{(x_k)}_{k}$ on the $\{2,\dots,N-1\}$ remaining qubits, we have 
\begin{eqnarray}
\mathbb{I}_2 \otimes \left( \bigotimes_{k=2}^N \Pi_k^{(x_k)} \right) \,\rho_\phi\, \mathbb{I}_2 \otimes \left( \bigotimes_{k=2}^N \Pi_k^{(x_k)} \right)%
&=& \bigg( \frac{\eta}{2} \ket{G_\phi^\mu}\bra{G_\phi^\mu}+ (1-\eta) \frac{\mathbb{I}_2}{2} \bigg) \nonumber \\
&& \otimes \left( \bigotimes_{k=2}^N \Pi_k^{(x_k)} \right) \times \frac{1}{2^{N-1}},
\end{eqnarray}
where $\ket{G_\phi^\mu} = \left(\ket{0} +(-1)^\mu e^{iN\phi}\ket{1} \right)/\sqrt{2}$. At this stage the probe is an uncorrelated state and all the information about $\phi$ was transferred to the qubit $1$. 

The last step is to manifest the phase in a probability distribution via the classically conditioned operation $\Pi^{(x_1)}_{1|2 \dots N}=(\sigma_z)^\mu H \ket{x_1}\bra{x_1} H (\sigma_z)^\mu$ on the qubit $X_1$. This operation removes the extra phase $(-1)^\mu$ from $\ket{G_\phi^\mu}\bra{G_\phi^\mu}$ resulting in 
\begin{equation*}
\bigg[ \eta \ket{\phi}\bra{\phi}+ (1-\eta)\frac{\mathbb{I}_2}{2} \bigg] \otimes \left( \bigotimes_{k=2}^N \Pi_k^{(x_k)} \right) \times \frac{1}{2^{N-1}},
\end{equation*}
where $\ket{\phi} = \cos\left(N\frac{\phi}{2}\right)\ket{0} + i\sin\left(N\frac{\phi}{2}\right)\ket{1}$. The Fisher information turns out to be 
\begin{equation}
\mathcal{F}_{\mbox{ad}}  = \sum_{k=2}^N \mathcal{F}(p_k) + \sum_{\{x_k\}} p_2(x_2) \dots p_N(x_N) \mathcal{F} (p_{1|x_2 \dots x_N}),
\end{equation}
with $\mathcal{F}(p_k)$ being the Fisher information for the local probability distribution, $p_k(x_k)$, associated to the measurement of $\Pi^{(x_k)}_k$ (on the $k$th qubit). Regarding the local distributions, we have $\partial_\phi p_k = 0$ for $k \neq 1$ implying that the local Fisher information vanishes, $\sum_{k=2}^N \mathcal{F}(p_k) = 0$. Moreover, the product $p_2(x_2) \times \dots \times p_N(x_N) = 1/2^{N-1}$ for any combination of outcomes $\{x_k ~|~ k=2,\dots,N \}$. 

The Fisher information of the multipartite adaptive strategy reduces to $\mathcal{F}_{\mbox{ad}} = \mathcal{F} (p_1)$, where $p_1 (x_1) = {\rm Tr~} \big( \rho_1 \ket{x_1}\bra{x_1} \big)$ (with $x_1=0,1$) and $\rho_1 = \bigg[ \eta \ket{\phi}\bra{\phi} + (1-\eta)\mathbb{I}_2/2 \bigg]$. Now, we have $p_\phi(0) = \eta\cos^2\left( N \frac{\phi}{2} \right) + \frac{1-\eta}{2}$ and $p_\phi(1) = \eta\sin^2\left( N \frac{\phi}{2} \right) + \frac{1-\eta}{2}$, and thus we have
\begin{equation}
\max_{\{\phi\}}\mathcal{F}_{\mbox{ad}} = N^2 \eta^2.
\label{POVMadpt}
\end{equation}

We observe that the maxima in Eqs. (\ref{POVMadpt}) and (\ref{POVMcoher}) occur at the same value, $\phi=\pi/(2N)$. Moreover, the Fisher informations in Eqs. (\ref{POVMadpt}) and (\ref{POVMcoher}) that comes from the POVM implemented in circuit depicted in Fig. \ref{StratFigMP}  are equal to the ones obtained from the QFI formula in the \ref{App2}. This shows that the POVMs implemented here are optimal at $\phi=\pi/(2N)$. It is worthwhile to observe that, we are interested in small deviations of a given value of phase. Since we only measure relative phases, we can always calibrate the interferometer to measure a small deviation from $\pi/(2N)$. So, the optimal POVM presented above can be used to measure a small deviation around any value of $\phi$ by suitably calibrating the measurement apparatus in our example. This of course assumes that $\mathcal{F}(\phi)$ is an analytic function, which is trivially true here; it is a mild assumption that can be expected to break down in only the most pathological examples.

%%%%%%%%%%%%%%%%%%%%%%%%%%%%%%%%%%%%%%%%%%%%%%%%%%%%%%%%%%%%
%%%%%%%%%%%%%%%%%%%%%%%%%%%%%%%%%%%%%%%%%%%%%%%%%%%%%%%%%%%%
\section{Existence of globally optimal measurements}\label{App4}
%%%%%%%%%%%%%%%%%%%%%%%%%%%%%%%%%%%%%%%%%%%%%%%%%%%%%%%%%%%%
%%%%%%%%%%%%%%%%%%%%%%%%%%%%%%%%%%%%%%%%%%%%%%%%%%%%%%%%%%%%
Here, we outline the proof of the result that full-rank density operators do not admit globally optimal measurements (that is, for any value of $\phi$); we shall begin by pointing out why this result of seemingly narrow scope is of wide applicability. Let $\mathcal{P(H)}$ denote the convex hull of all density operators $\{\varrho \in \mathcal{P(H)} : \varrho \geq 0,\tr(\varrho) = 1\}$ on a finite-dimensional Hilbert space $\mathcal{H}$ where $d = \mathrm{dim}(\mathcal{H})$. $\mathcal{P(H)}$ can be naturally partitioned into subsets of given rank:
\begin{equation}
\mathcal{P}(\mathcal{H}) = \bigcup_{j=1}^{d} \mathcal{P}_{j} \; \mbox{where} \; \{\varrho \in \mathcal{P}_{j}: \mathrm{rank}(\varrho) = j \}.
\end{equation}
Now suppose we take a rank-deficient state ($\varrho \in \mathcal{P} \setminus \mathcal{P}_d$) and add an arbitrarily small amount of noise, we obtain the state $\tilde{\varrho} = \eta \frac{\mathbb{I}}{d} + (1-\eta)\varrho$. Exploiting the fact that density operators are Hermitian and therefore unitarily diagonalisable, we can re-write this in the form $\tilde{\varrho} = V^\dagger (\eta \frac{\mathbb{I}}{d} + (1-\eta)D) V$, where $D$ is the diagonal representation of $\varrho$, we can see immediately that the noise has perturbed all the eigenvalues, causing the state $\tilde{\varrho}$ to collapse into the subset $P_d$. Since noise is inevitable, the aforementioned result applies to any practically realisable state. We shall now turn to the proof.

Consider a state $\varrho_\phi$ that is full-rank for all $\phi$ and impose the further mild assumption that the symmetric logarithmic derivative $L_\phi$ is non-degenerate (all eigenvalues are distinct) at least one value of $\phi$. Using the result that any POVM can be fine-grained to rank-one elements (without reduction of the Fisher information), Barndorff-Nielsen et al.~\cite{Barndorff-Nielsen2} have shown that any POVM on $\varrho_\phi$ satisfying Eq.~(\ref{opt-povm-conds}) can be fine-grained to a projective measurement of $L_\phi$ (discussed above). It now remains to establish the condition for the observable $L_\phi$ to be $\phi$-independent.

The starting point is a result of Nagaoka: if a globally optimal measurement exists, which we shall assume to be fine-grained to an observable $T$, then $\varrho_{\phi}$ is a member of the quantum exponential family (Barndorff-Nielsen et al. helpfully recapitulate this in detail~\cite{Barndorff-Nielsen2}). Fujiwara has used information geometry to develop an equivalent condition: $\varrho_\phi$ must be an e-geodesic on the Riemannian manifold of quantum states~\cite{Hayashi}. In our case, we have unitary families of states $\varrho_{\phi} = e^{i \phi H} \varrho e^{-i \phi H}$ for some Hermitian generator $H$; Fujiwara has proved that $\varrho_\phi$ is an e-geodesic if and only if $[H,\mathcal{D}_0\, H] = i \mathcal{F}(\varrho_{\phi})$, where $\mathcal{D}_0$ is the rather unusual commutation operator $\mathcal{D}$ (at $\phi = 0$): $i[X,\varrho] = \frac{1}{2} \{\mathcal{D} X, \varrho\}$~\cite{Fujiwara}.

It turns out that we can readily recast the condition in terms of a more familiar operator as $[L_0,H] = i \fisher(\varrho_{\phi})$, where $L_0$ is the SLD in the Heisenberg picture $L_0 = U_\phi^\dagger L_{\phi} U_\phi$. However, it is a result of functional analysis that $[A,B] \propto i$ (canonical commutation relation) cannot be satisfied in finite-dimensional Hilbert spaces -- this trivially follows from taking the trace (well-defined for finite $d$) of the commutation relation and noting that the left-hand side vanishes by virtue of the cyclic invariance of the trace, whereas the right-hand side is proportional to the trace of the identity. Since we are exclusively interested in finite $d$, this completes the proof.

%%%%%%%%%%%%%%%%%%%%%%%%%%%%%%%%%%%%%%%%%%%%%%%%%%%%%%%%%%%%
%%%%%%%%%%%%%%%%%%%%%%%%%%%%%%%%%%%%%%%%%%%%%%%%%%%%%%%%%%%%

\end{document}